\documentclass[pre,twocolumn,showpacs]{revtex4}
\usepackage{epsfig}
\usepackage{amsmath}
\usepackage{amssymb}

\begin{document}

\title{Hamilton's equations of motion of a vortex filament
in the rotating Bose-Einstein condensate
and their ``soliton'' solutions}
\author{V. P. Ruban}
\email{ruban@itp.ac.ru}
\affiliation{Landau Institute for Theoretical Physics RAS, Moscow, Russia} 

\date{\today}

\begin{abstract}
The equation of motion of a quantized vortex filament in a trapped Bose-Einstein condensate 
[A. A. Svidzinsky and A. L. Fetter, Phys. Rev. A {\bf 62}, 063617 (2000)] has been generalized to the 
case of an arbitrary anharmonic anisotropic rotating trap and presented in a variational form. 
For condensate density profiles of the form  $\rho=f(x^2+y^2+\mbox{Re\,}\Psi(x+iy))$
in the presence of the plane of symmetry $y=0$, the solutions $x(z)$ describing
stationary vortices of U and S types coming to the surface and solitary waves have been found in quadratures.
Analogous three-dimensional configurations of the vortex filament uniformly moving along the $z$ axis have
also been found in strictly cylindrical geometry. The dependence of solutions on the form of the function $f(q)$
has been analyzed.
\end{abstract}

%03.75.Kk Dynamic properties of condensates; collective and hydrodynamic excitations, 
%superfluid flow
%67.85.De Dynamic properties of condensates; excitations, and superfluid flow

\maketitle

{\bf Introduction}.
The dynamics of quantized vortex filaments is an
important part of Bose condensed gas physics (see,
e.g., reviews [1, 2] and numerous references therein).
At low temperatures, a Bose-Einstein condensate is
described well by the Gross-Pitaevskii equation
\begin{equation}
i\hslash\psi_t=-\frac{\hslash^2}{2m}\Delta\psi+ V({\bf r}_{\rm lab},t)\psi +g|\psi|^2\psi.
\end{equation}
At rather fast rotation of the trap potential $V({\bf r}_{\rm lab},t)$, 
the vortex-free state of the quantum gas of atoms with
the mass $m$  is energetically unfavorable and the vortex
with the circulation $\Gamma= 2\pi\hslash/m$ penetrates inside the
condensate. The vortex filament with its own velocity field
exists against the background of the nonuniform
unperturbed (mass) density $\rho_0=m|\psi_0|^2$  and the 
unperturbed velocity ${\bf v}_0$, which are time-independent in the
coordinate system rotating about the $z$ axis: 
\begin{equation}\label{stat}
{\bf v}_0=\Omega(\nabla \varphi-[{\bf e}_z\times {\bf r}]), 
\qquad \nabla\cdot(\rho_0 {\bf v}_0)=0.
\end{equation}
We assume that the rotation frequency $\Omega$ is much
lower than the characteristic transverse frequency of
the trap $\omega_\perp$, which provides the stability of the unperturbed 
flow [3, 4]. Then, the density field $\rho_0({\bf r})$ in the
Thomas-Fermi approximation and ignoring the
terms of order $(\Omega/\omega_\perp)^2$ is determined from the equation
\begin{equation}
V({\bf r})+ (g/m)\rho_0({\bf r}) \approx \mu=const,
\end{equation}
where $V({\bf r})$ is the trap potential in the rotating system
and $\mu$ is the rather large chemical potential of the gas
satisfying the condition $\mu\gg \hslash\omega_\perp$. We note that the
characteristic transverse size of the condensate is $R_0\sim(\mu/m\omega_\perp^2)^{1/2}$, 
while the characteristic thickness of the
vortex core is $\xi\sim\hslash/(m\mu)^{1/2}$. Their ratio is large, 
$R_0/\xi \sim \mu/\hslash \omega_\perp \gg 1$, which makes it possible to use the 
so-called local induction approximation in the dynamics
of the vortex filament. The condition $\rho_0({\bf r})=0$ 
determines the (usually closed) Thomas-Fermi surface $\Sigma$,
which can be considered as an effective boundary of
the condensate. In practice, the topology of the surface  $\Sigma$
is most often spherical or toroidal (see, e.g., [5, 6]) 
but also can be more complicated as a function of
the trap potential: the condensate may have several
through “holes” and/or internal cavities.

Since the vortex filament is transported by the total
velocity field, its equation of motion in the local
induction approximation has the form
\begin{equation}\label{LIA}
{\bf R}_t =\frac{\Gamma\Lambda}{4\pi} \{\kappa {\bf b} 
+[(\nabla \rho_0({\bf R})/\rho_0({\bf R}))\times {\bf t}]\}
+{\bf v}_0({\bf R}),
\end{equation}
where $\Lambda=\log(R_0/\xi)\approx \log(\mu/\hslash\omega_\perp) \approx const \gg 1$ 
is the
large logarithm, $\kappa$  is the local curvature of the filament, 
${\bf t}$ is the tangent unit vector, and  ${\bf b}$ is the binormal
unit vector. For the harmonic trap, this equation was
derived in [7] using crosslinked asymptotic expansions
and then it was used in the linearized form in several
works [8, 9]. It is significant that the Hamiltonian
structure of Eq. (4) was not discussed earlier except for
the case $\Omega =0$ (see [10], where the local induction
equation against the static inhomogeneous background was derived from the 
noncanonical Hamiltonian formalism). However, it is known that the 
stationary solutions of Eq. (4) in the case of the quadratic
potential are extremals of a certain functional [11].

This work is devoted to the formulation of a variational 
principle for Eq. (4) in the general case and the
determination of analytical nonlinear solutions for
some (rather broad) class of density profiles uniform
over $z$. Although real Bose–Einstein condensates are
limited in the axial direction, the analytical solutions
obtained may be approximately correct for rather long
condensates.

{\bf Variational formulation}.
Let the position of the vortex filament in space be given
by a vector function ${\bf R}(\beta, t)$ with an arbitrary 
longitudinal parameter $\beta$. We note that the existence of a
vector potential for unperturbed current density follows from Eq. (2):
\begin{equation}
\rho_0(\nabla \varphi-[{\bf e}_z\times {\bf r}])= \mbox{\,curl\,}{\bf A}.
\end{equation}
The vector potential ${\bf A}({\bf r})$ satisfies the equation
\begin{equation}\label{A_eq}
\mbox{curl\,}\frac{1}{\rho_0({\bf r})} \mbox{curl\,}{\bf A}=-2{\bf e}_z.
\end{equation}
It is easy to check that Eq. (4) can be rewritten as fol-
lows (see details in [10] for the case $\Omega =0$):
\begin{equation}\label{variat}
\Gamma\,[{\bf R}_\beta \times{\bf R}_t]\rho_0({\bf R})=\delta{\cal H}/\delta{\bf R},
\end{equation}
where the Hamiltonian functional is
\begin{equation}
{\cal H}=\frac{\Gamma^2}{4\pi}\Lambda \int \rho_0({\bf R})|{\bf R}_\beta|d\beta 
+\Gamma\Omega\int ({\bf A}({\bf R})\cdot {\bf R}_\beta) d \beta.
\end{equation}
Under the boundary condition ${\bf A}|_{\Sigma}=0$ , this functional
is approximately (owing to the relation $R_0/\xi\gg 1$) the
part of the total kinetic energy of the flow that depends
on the position of the vortex filament. It is important
that Eq. (7) follows from the variational principle with
the Lagrangian
\begin{equation}
{\cal L}=\Gamma \int({\bf D}({\bf R})\cdot[{\bf R}_\beta \times{\bf R}_t]) d\beta -{\cal H},
\end{equation}
and the vector function ${\bf D}({\bf R})$ satisfies the equation
\begin{equation}
\nabla\cdot{\bf D}({\bf R})=\rho_0({\bf R}).
\end{equation}

Thus, the required variational formulation of the
dynamics of a single vortex filament is found in terms of
the vector potential ${\bf A}$. The generalization for several
filaments is formally simple but is technically difficult,
since it is necessary to take into account the nonlocal
interaction between filaments; to this end, it is 
necessary to know the solution of the equation
\begin{equation}
\mbox{curl\,}\frac{1}{\rho_0({\bf r})} \mbox{curl\,}{\bf A}_\Gamma={\bf\Omega}_\Gamma,
\end{equation}
where ${\bf\Omega}_\Gamma$ is the singular vorticity field created by 
filaments and ${\bf A}_\Gamma$ is the vector potential of the 
corresponding current density.

If necessary, using results of [12], a variational
principle in the dynamics of the vortex filament can
also be formulated in the case of a completely 
time-dependent background density profile. This means
that the trap potential not only rotates but also is
deformed. We do not discuss this problem here.

{\bf Case of the global stream function}.
For the further more particular consideration, it is
necessary to have the explicit solution of Eq. (6). We
consider the case where the unperturbed velocity field
is two-dimensional (independent of the coordinate $z$)
and solenoidal. This means that there exists the stream
function of the unperturbed velocity common for all
$z$ values:
\begin{equation}
\Theta(x,y)=\frac{\Omega}{2}[x^2+y^2+\theta(x,y)],\quad \theta_{xx}+\theta_{yy}=0,
\end{equation}
where the harmonic function $\theta(x,y)$ determines the
axial asymmetry of the flow. The density field in this
case has the form
\begin{equation}
\rho_0({\bf r})= f(x^2+y^2+\theta(x,y); z),
\end{equation}
where $f(q;z)$ is a rather arbitrary function of two
variables. We introduce the function $Q(z)$ determined
by the condition $f(Q;z)=0$ and the function
\begin{equation}
 F(q;z)=\int_q^{Q(z)} f(u;z)du,
\end{equation}
which is positive inside the condensate and is zero at
its boundary. Then, the vector potential is
\begin{equation}
{\bf A}({\bf r})=-\frac{{\bf e}_z}{2} F(x^2+y^2+\theta(x,y); z). 
\end{equation}

If the condensate has no holes, the harmonic function $\theta(x,y)$  
can be represented in the form of the sum
\begin{equation}
\theta(x,y)= \mbox{Re\,}\sum_{n\ge 2} c_n (x+iy)^n
\end{equation}
with arbitrary complex coefficients $c_n$. If the surface $\Sigma$
is not topologically equivalent to a sphere, the diversity
of analytical functions of the complex variable  $\Psi(x+iy)$
makes it possible to consider rather exotic forms of
condensates, including those with the quantized circulations 
of the unperturbed velocity $l_j \Gamma$ around each hole.

It should be noted that the quadratic anisotropic
trap corresponds to the choice $\theta=\epsilon (x^2-y^2)$ and
$$
f_{quadr}\propto 1-(1+\epsilon)x^2-(1-\epsilon)y^2 -z^2/z_{max}^2.
$$
Here and below, all lengths are given in units of $R_0$ and
all times are measured in units of $2\pi R_0^2/\Gamma\Lambda\gg 1/\omega_\perp$.

We choose the coordinate $z$ as the longitudinal
parameter $\beta$. Then, ${\bf R}=(x(z,t), y(z,t), z)$ and the
dimensionless Hamiltonian is written in the form
\begin{eqnarray}
H&=&\frac{1}{2}\int f(x^2+y^2+\theta(x,y); z)\sqrt{1+x_z^2+y_z^2} dz -\nonumber \\ 
&&-\frac{\tilde\Omega}{2}\int F(x^2+y^2+\theta(x,y); z)dz,
\end{eqnarray}
where $\tilde\Omega=2\pi R_0^2\Omega/\Gamma\Lambda$.
The corresponding noncanonical Hamilton’s equations of motion have the structure
\begin{equation}
 x_t =\frac{1}{f}\frac{\delta H}{\delta y},\qquad
-y_t =\frac{1}{f}\frac{\delta H}{\delta x}.
\end{equation}

{\bf Analytical solutions}.
Stationary (in the rotating coordinate system) configurations of the vortex filament are extremals of the
functional $H$. In a number of cases, they can be calculated analytically, e.g., when $f$ is independent of $z$
and all coefficients $c_n$ are real. The vortex filament lies in the plane of symmetry $y=0$, and the minimizing
functional has the form
\begin{equation}
2\hat H=\int f(\alpha(x))\sqrt{z_\beta^2+x_\beta^2} d\beta
-\tilde\Omega\int F(\alpha(x))z_\beta d\beta,
\end{equation}
where $\alpha(x)=x^2+\sum_{n\ge 2} c_n x^n$. The corresponding
Euler-Lagrange equations are integrated:
\begin{equation}
f\frac{dz}{ds}-\tilde\Omega F=E=const, 
\end{equation}
\begin{equation}
\left(\frac{dx}{ds}\right)^2=1-\left(\frac{E+\tilde\Omega F}{f}\right)^2,
\end{equation}
where  $ds$ is the arc length element. In the case of zero
integration constant  $E$, the so-called U and S vortices
coming on the surface are obtained (see, e.g., [13-15]). If  $E=f(0)-\tilde\Omega F(0)$,
the solutions are soliton-like
dependences $x(z)$; in some cases, these can be curves
with self-crossing. Self-crossing solutions certainly
violate the condition of applicability of the local
induction approximation and, therefore, they are not
physical. Corresponding examples are given in Figs. 1
and 2 for the simplest choice 
$q=(1+\epsilon)x^2+(1-\epsilon)y^2$, 
$\epsilon =9/16$, $f(q)=1-q$.

\begin{figure}
\begin{center}
   \epsfig{file=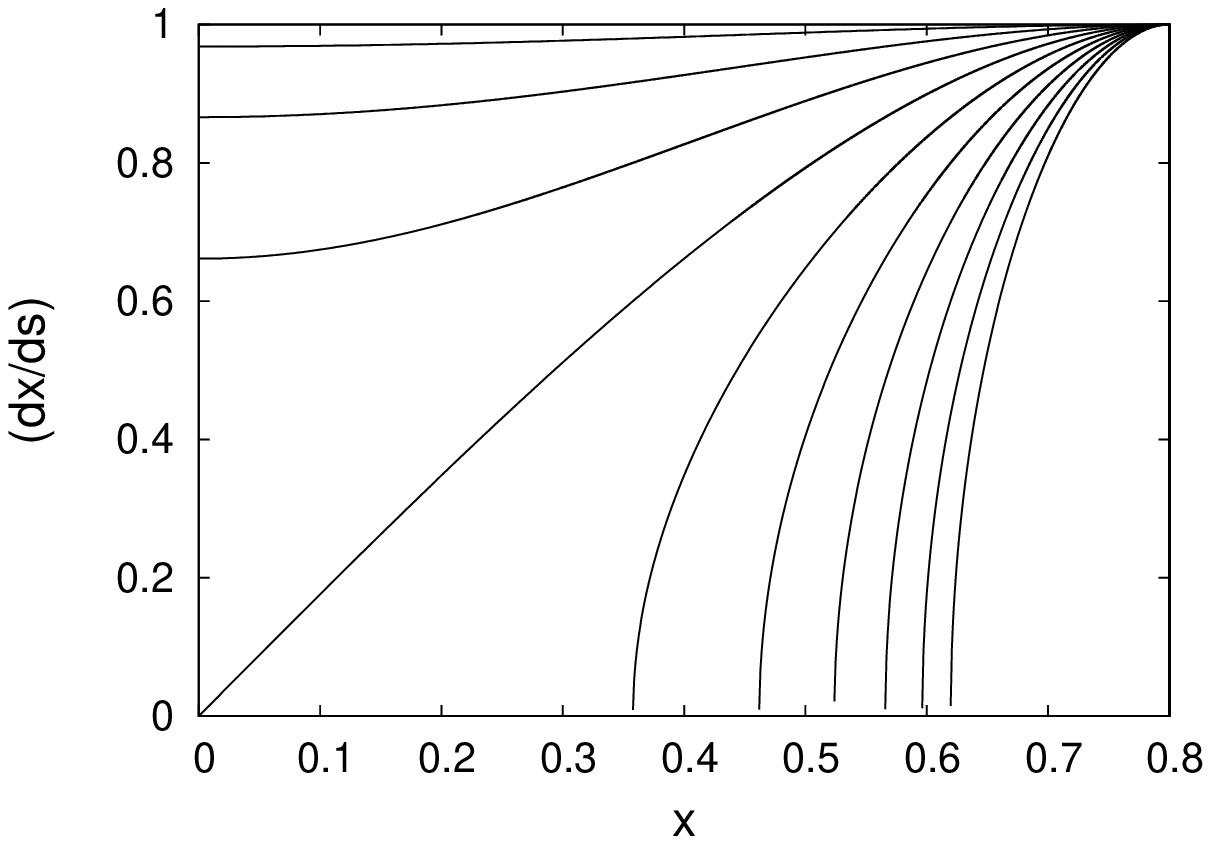, width=80mm}\\
\vspace{3mm}
   \epsfig{file=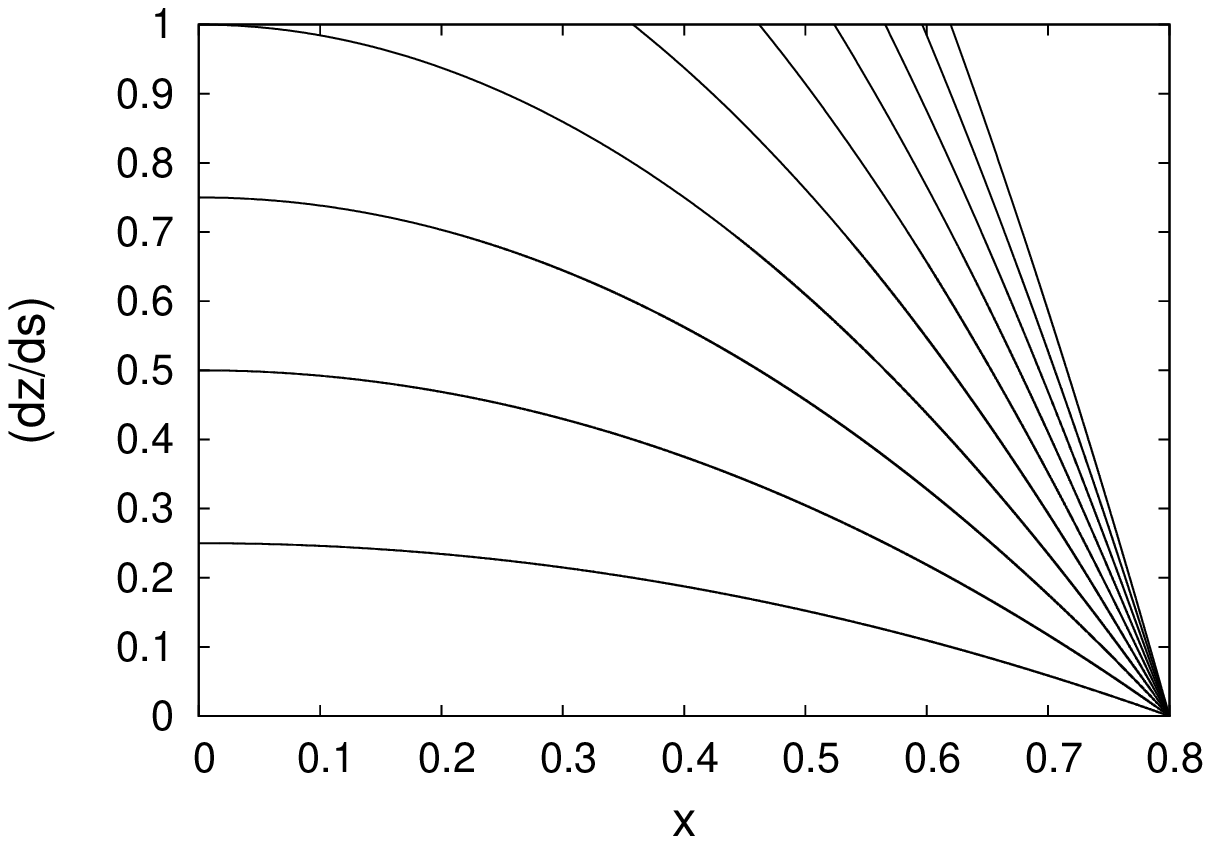, width=80mm}
\end{center}
\caption{
Examples of phase curves for U and S vortices at different $\tilde\Omega$ values.
At $\tilde\Omega<2$, S vortices occur (reaching the value  $x=0$); At $\tilde\Omega>2$, 
U vortices exist. At $\tilde\Omega=2$, half the
limiting soliton without self-crossing is obtained.
}
\label{U-S-vortices} 
\end{figure}

\begin{figure}
\begin{center}
   \epsfig{file=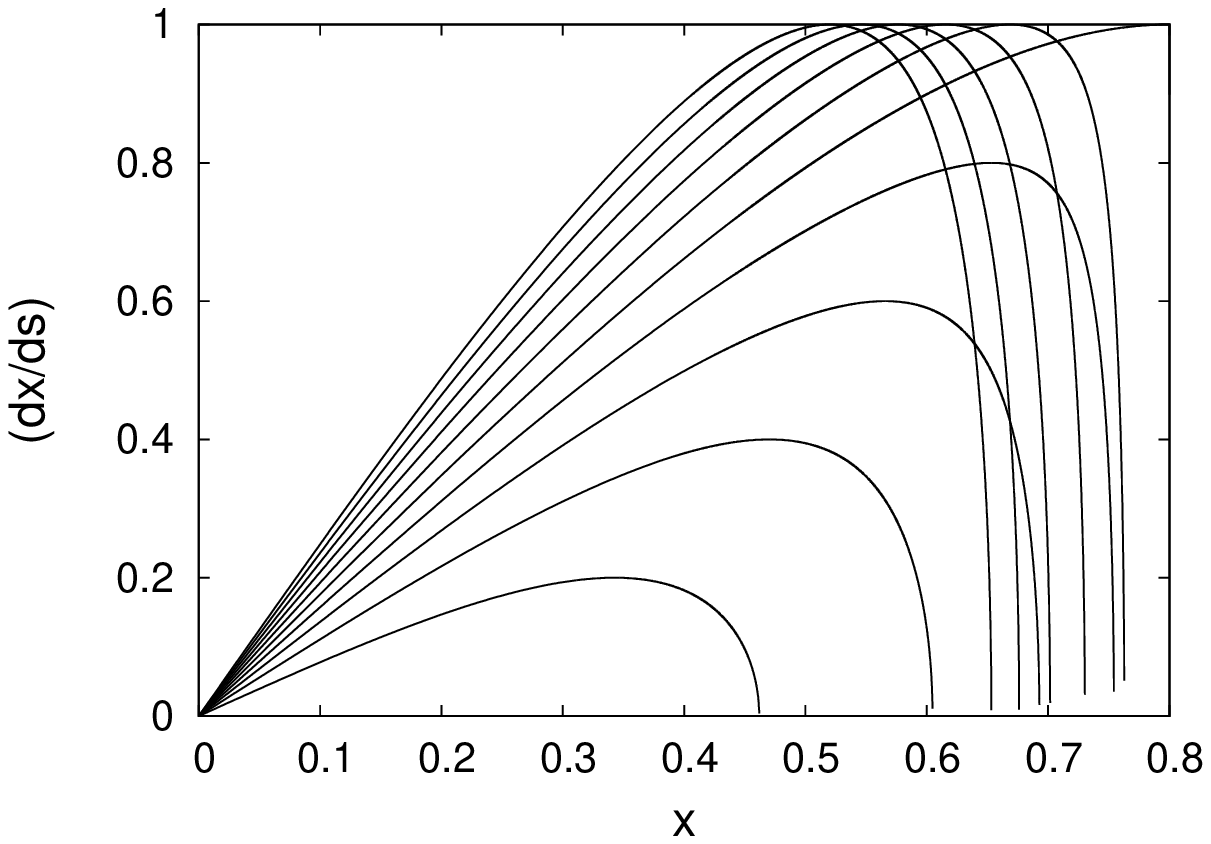, width=80mm}\\
\vspace{3mm}
   \epsfig{file=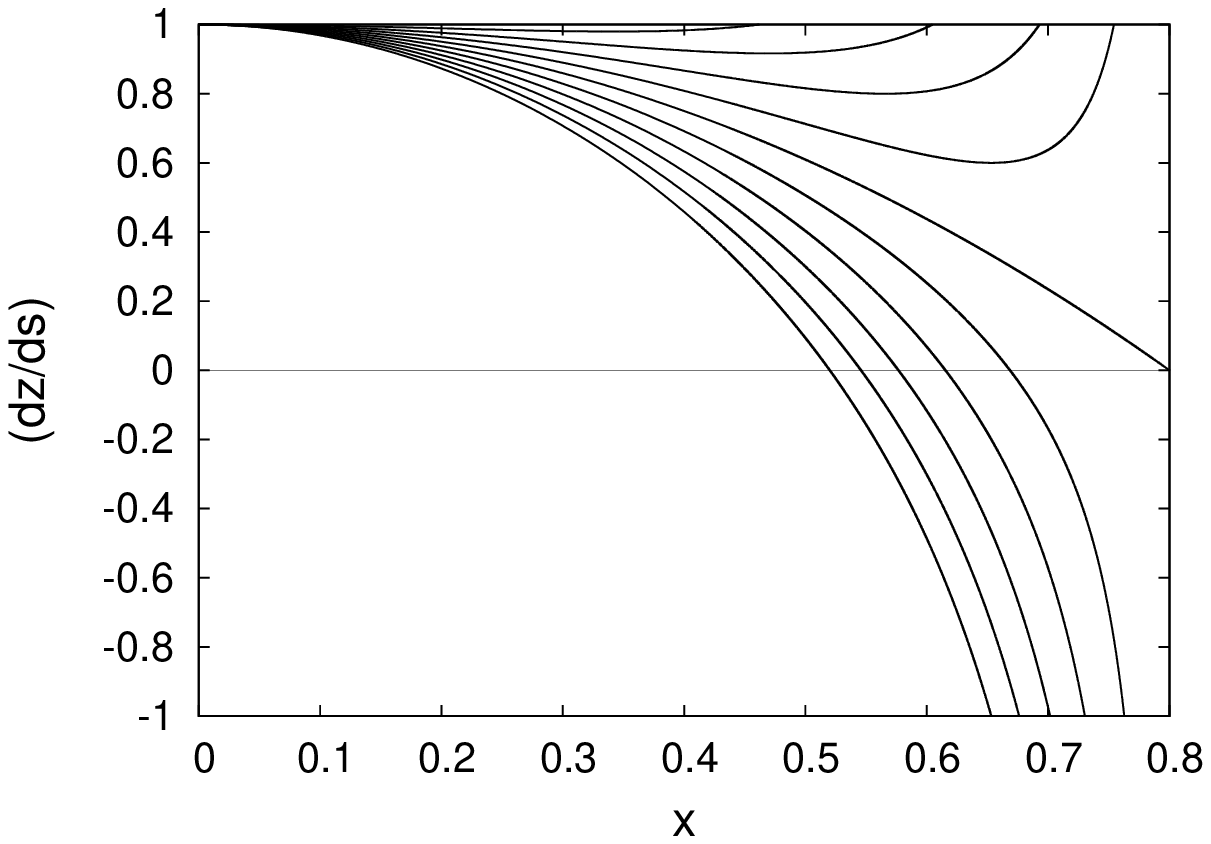, width=80mm}
\end{center}
\caption{
Examples of phase curves for solitons at different $\tilde\Omega$ values. 
At  $1<\tilde\Omega<2$, solutions are ``good''; at $\tilde\Omega>2$, curves
with self-crossing are obtained.}
\label{solitons} 
\end{figure}

{\bf Cylindrical geometry}.
The solution of the problem is the most advanced
in the presence of cylindrical symmetry. The density
profile is static in the laboratory coordinate system
and it is not necessary to transfer to the rotating system
for the formulation of the variational principle. It is
convenient to introduce the complex unknown variable  
$w(z,t)=r(z,t)e^{i\phi(z,t)}=x(z,t)+iy(z,t)$. Now, the
vector function ${\bf D}({\bf R})$ has the explicit form
\begin{equation}
{\bf D}({\bf R})=\frac{G(r^2)}{2r}(\cos\phi, \sin \phi, 0), 
\end{equation}
where
\begin{equation}
G(q)=\int_0^q f(u) du = F(0) - F(q).
\end{equation}
is a nonnegative function. The Lagrangian of the filament is determined by the expression
\begin{eqnarray}
L_*&=&\int \frac{i}{2} G(|w|^2)\left[\frac{w_t}{w}-\frac{w^*_t}{w^*}\right] d z -\nonumber\\
&&-\int f(|w|^2)\sqrt{1+|w_z|^2} dz,
\end{eqnarray}
and its equation of motion has the form
\begin{eqnarray}
iw_t&=&\frac{f'(|w|^2)}{f(|w|^2)}\left[\frac{2w-w_z^2w^*+|w_z|^2w}{2\sqrt{1+|w_z|^2}}\right]-
\nonumber\\
&&-\frac{\partial}{\partial z}\left(\frac{w_z}{2\sqrt{1+|w_z|^2}}\right).
\end{eqnarray}
Integrals of motion of this equation (in addition to the Hamiltonian
$H_*=\int f(|w|^2)\sqrt{1+|w_z|^2}\,dz$) are given by the formulas
\begin{eqnarray}
N&=&\int G(|w|^2)dz = const,\\ 
P&=&-\int\frac{i}{2} G(|w|^2)\left[\frac{w_z}{w}-\frac{w^*_z}{w^*}\right] dz = const.
\end{eqnarray}

Solutions of the form $w=r(z-vt)\exp[i\phi(z-vt)+i\tilde\Omega t]$
give the extremum to the functional 
\begin{equation}
\tilde H_*=H_* + \tilde\Omega N-vP =\int \left(f ds + \tilde\Omega G d z - v G d\phi \right),
\end{equation}
where $ ds=\sqrt{dz^2+dr^2+r^2d\phi^2}$.  The formulas below
follow from the requirement  $\delta\tilde H_*=0$ 
\begin{equation}
\frac{d\phi}{ds}=\frac{vG+M}{r^2f}, \qquad  \frac{dz}{ds}=\frac{C-\tilde\Omega G}{f},
\end{equation}
with constant $M$ and $C$. Then, we obtain
\begin{equation}
\left(\frac{dr}{ds}\right)^2=
\frac{f^2(r^2)-[vG(r^2)+M]^2/r^2-[C-\tilde\Omega G(r^2)]^2}{f^2(r^2)}.
\end{equation}
It is noteworthy that the curve at$v\not = 0$ or at $M\not = 0$ is
necessarily three-dimensional. For three-dimensional
``solitons,'' the integration constants are $M=0$ and $C=f(0)$.
For three-dimensional U and S vortices, $M=-vG(r^2_{max})$ and $C=\tilde\Omega G(r^2_{max})$. 

The small-angle approximation $|w_z|^2\ll 1$ deserves
special attention, since it allows qualitative analysis of
the dependence of the dynamics on the density profile. 
In this case, the Hamiltonian and equation of
motion are given by the formulas 
\begin{equation}
{\cal H}_{(2)}=\int f(|w|^2)\left(1+\frac{|w_z|^2}{2}\right) dz,
\end{equation}
\begin{equation}\label{NLSE_general}
iw_t=-\frac{1}{2}w_{zz}+\frac{f'(|w|^2)}{f(|w|^2)}\left[w-\frac{w_z^2w^*}{2}\right].
\end{equation}
At small $|w|^2$ values, we have $f'(|w|^2)/f(|w|^2)\approx c_0+c_1|w|^2$.
If $c_1<0$ and $c_1>0$, Eq. (32) in the long-wavelength
limit is similar to the focusing and defocusing nonlinear Schrödinger equations, 
respectively. We also note that the Gaussian density profile $f(r^2)=\exp(-r^2)$ gives
a quite compact nonlinear Schrödinger-type equation
but with another cubic nonlinearity:
\begin{equation}
iw_t=-\frac{1}{2}w_{zz}-w+\frac{w_z^2w^*}{2}.
\end{equation}

{\bf Conclusions}.
Thus, the variational formulation of the dynamics
of the vortex filament in the rotating Bose-Einstein
condensate made it possible to find analytically some
stationary solutions. In particular, in addition to the
known vortices of the S and U types, the soliton-like
configurations, which apparently have not yet been
discussed, have been obtained. Noncanonical 
Hamiltonian equations of motion derived in this work can be
used further for the numerical simulation of the
dynamics of the filament. The interaction of solitons
can possibly lead to the phenomenon of anomalous
waves (rogue waves) and be accompanied by the
approach of the vortex filament on the surface with its
subsequent rupture.

\end{document}